\documentclass[aps,prl,10pt,twocolumn,superscriptaddress,amsmath,amssymb]{revtex4-2}

\usepackage{graphicx} % Include figure files
\usepackage{dcolumn}  % Align table columns on decimal point
\usepackage{bm}       % bold math
\usepackage{multirow}
\usepackage{color}
\usepackage{comment}
\usepackage{booktabs}
\usepackage{orcidlink} % For ORCID links in author list

\usepackage{hyperref} % color should be defined below

\usepackage{tensor}
\usepackage{tikz}
\usepackage{dsfont}

\definecolor{kblue}{RGB}{20,135,200}
\definecolor{bpppink}{RGB}{255,89,199}
\definecolor{chv3lblue}{HTML}{22CCF2}
\definecolor{blueroseblue}{HTML}{7481D2}
\definecolor{tpgreen}{HTML}{324810}

\hypersetup{
  colorlinks = true,    % turn on colored text for links
  linkcolor = kblue,    % internal cross‐references
  citecolor  = bpppink,   % citations
  urlcolor   = blueroseblue     % URLs (was magenta by default)
}

\begin{document}

\title{Probing unexplored spin-dependent dark matter--proton coupling with few-photoelectron threshold in COSINE-100}

\author{W.~K.~Kim}
\email{wonkkim30@gmail.com}
\affiliation{IBS School, University of Science and Technology (UST), Daejeon 34113, Republic of Korea}
\affiliation{Center for Underground Physics, Institute for Basic Science (IBS), Daejeon 34126, Republic of Korea}
\author{N.~Carlin}
\affiliation{Physics Institute, University of S\~{a}o Paulo, 05508-090, S\~{a}o Paulo, Brazil}
\author{J.~Y.~Cho}
\affiliation{Department of Physics, Kyungpook National University, Daegu 41566, Republic of Korea}
\author{S.~J.~Cho}
\affiliation{Center for Underground Physics, Institute for Basic Science (IBS), Daejeon 34126, Republic of Korea}
\author{S.~Choi}
\affiliation{Department of Physics and Astronomy, Seoul National University, Seoul 08826, Republic of Korea} 
\author{A.~C.~Ezeribe}
\affiliation{Department of Physics and Astronomy, University of Sheffield, Sheffield S3 7RH, United Kingdom}
\author{L.~E.~Fran{\c c}a}
\affiliation{Physics Institute, University of S\~{a}o Paulo, 05508-090, S\~{a}o Paulo, Brazil}
\author{R.~F.~Muhdi}
\affiliation{Department of Physics, Universitas Negeri Malang, Malang 65145, Indonesia}
\author{O.~Gileva}
\affiliation{Center for Underground Physics, Institute for Basic Science (IBS), Daejeon 34126, Republic of Korea}
\author{C.~Ha}
\affiliation{Department of Physics, Chung-Ang University, Seoul 06973, Republic of Korea}
\author{I.~S.~Hahn}
\affiliation{Center for Exotic Nuclear Studies, Institute for Basic Science (IBS), Daejeon 34126, Republic of Korea}
\affiliation{Department of Science Education, Ewha Womans University, Seoul 03760, Republic of Korea} 
\affiliation{IBS School, University of Science and Technology (UST), Daejeon 34113, Republic of Korea}
\author{E.~J.~Jeon}
\affiliation{Center for Underground Physics, Institute for Basic Science (IBS), Daejeon 34126, Republic of Korea}
\affiliation{IBS School, University of Science and Technology (UST), Daejeon 34113, Republic of Korea}
\author{H.~W.~Joo}
\affiliation{Department of Physics and Astronomy, Seoul National University, Seoul 08826, Republic of Korea} 
\author{W.~G.~Kang}
\affiliation{Center for Underground Physics, Institute for Basic Science (IBS), Daejeon 34126, Republic of Korea}
\author{M.~Kauer}
\affiliation{Department of Physics and Wisconsin IceCube Particle Astrophysics Center, University of Wisconsin-Madison, Madison, WI 53706, USA}
\author{B.~H.~Kim}
\affiliation{Center for Underground Physics, Institute for Basic Science (IBS), Daejeon 34126, Republic of Korea}
\author{D.~Y.~Kim}
\affiliation{Center for Underground Physics, Institute for Basic Science (IBS), Daejeon 34126, Republic of Korea}
\author{H.~J.~Kim}
\affiliation{Department of Physics, Kyungpook National University, Daegu 41566, Republic of Korea}
\author{J.~Kim}
\affiliation{Department of Physics, Chung-Ang University, Seoul 06973, Republic of Korea}
\author{K.~W.~Kim}
\affiliation{Center for Underground Physics, Institute for Basic Science (IBS), Daejeon 34126, Republic of Korea}
\author{S.~H.~Kim}
\affiliation{Center for Underground Physics, Institute for Basic Science (IBS), Daejeon 34126, Republic of Korea}
\author{S.~K.~Kim}
\affiliation{Department of Physics and Astronomy, Seoul National University, Seoul 08826, Republic of Korea}
\author{Y.~D.~Kim}
\affiliation{Center for Underground Physics, Institute for Basic Science (IBS), Daejeon 34126, Republic of Korea}
\affiliation{IBS School, University of Science and Technology (UST), Daejeon 34113, Republic of Korea}
\author{Y.~H.~Kim}
\affiliation{Center for Underground Physics, Institute for Basic Science (IBS), Daejeon 34126, Republic of Korea}
\affiliation{IBS School, University of Science and Technology (UST), Daejeon 34113, Republic of Korea}
\author{B.~R.~Ko}
\affiliation{Department of Accelerator Science, Korea University, Sejong 30019, Republic of Korea}
\author{Y.~J.~Ko}
\email{yjkophys@jejunu.ac.kr}
\affiliation{Department of Physics, Jeju National University, Jeju 63243, Republic of Korea}
\author{B.~C.~Koh}
\affiliation{Department of Physics, Chung-Ang University, Seoul 06973, Republic of Korea}
\author{D.~H.~Lee}
\affiliation{Department of Physics, Kyungpook National University, Daegu 41566, Republic of Korea}
\author{E.~K.~Lee}
\affiliation{Center for Underground Physics, Institute for Basic Science (IBS), Daejeon 34126, Republic of Korea}
\author{H.~Lee}
\affiliation{IBS School, University of Science and Technology (UST), Daejeon 34113, Republic of Korea}
\affiliation{Center for Underground Physics, Institute for Basic Science (IBS), Daejeon 34126, Republic of Korea}
\author{H.~S.~Lee}
\email{hyunsulee@ibs.re.kr}
\affiliation{Center for Underground Physics, Institute for Basic Science (IBS), Daejeon 34126, Republic of Korea}
\affiliation{IBS School, University of Science and Technology (UST), Daejeon 34113, Republic of Korea}
\author{H.~Y.~Lee}
\affiliation{Center for Exotic Nuclear Studies, Institute for Basic Science (IBS), Daejeon 34126, Republic of Korea}
\author{I.~S.~Lee}
\affiliation{Center for Underground Physics, Institute for Basic Science (IBS), Daejeon 34126, Republic of Korea}
\author{J.~Lee}
\affiliation{Center for Underground Physics, Institute for Basic Science (IBS), Daejeon 34126, Republic of Korea}
\author{J.~Y.~Lee}
\affiliation{Department of Physics, Kyungpook National University, Daegu 41566, Republic of Korea}
\author{M.~H.~Lee}
\affiliation{Center for Underground Physics, Institute for Basic Science (IBS), Daejeon 34126, Republic of Korea}
\affiliation{IBS School, University of Science and Technology (UST), Daejeon 34113, Republic of Korea}
\author{S.~H.~Lee}
\affiliation{IBS School, University of Science and Technology (UST), Daejeon 34113, Republic of Korea}
\affiliation{Center for Underground Physics, Institute for Basic Science (IBS), Daejeon 34126, Republic of Korea}
\author{S.~H.~Lee}
\affiliation{Department of Physics, Jeju National University, Jeju 63243, Republic of Korea}
\author{Y.~J.~Lee}
\affiliation{Department of Physics, Chung-Ang University, Seoul 06973, Republic of Korea}
\author{D.~S.~Leonard}
\affiliation{Center for Underground Physics, Institute for Basic Science (IBS), Daejeon 34126, Republic of Korea}
\author{N.~T.~Luan}
\affiliation{Department of Physics, Kyungpook National University, Daegu 41566, Republic of Korea}
\author{V.~H.~A.~Machado}
\affiliation{Physics Institute, University of S\~{a}o Paulo, 05508-090, S\~{a}o Paulo, Brazil}
\author{B.~B.~Manzato}
\affiliation{Physics Institute, University of S\~{a}o Paulo, 05508-090, S\~{a}o Paulo, Brazil}
\author{R.~H.~Maruyama}
\affiliation{Department of Physics and Wright Laboratory, Yale University, New Haven, CT 06520, USA}
\author{S.~L.~Olsen}
\affiliation{Center for Underground Physics, Institute for Basic Science (IBS), Daejeon 34126, Republic of Korea}
\author{H.~K.~Park}
\affiliation{Department of Accelerator Science, Korea University, Sejong 30019, Republic of Korea}
\author{H.~S.~Park}
\affiliation{Korea Research Institute of Standards and Science, Daejeon 34113, Republic of Korea}
\author{J.~C.~Park}
\affiliation{Department of Physics and Institute for Sciences of the Universe, Chungnam National University, Daejeon 34134, Republic of Korea}
\author{J.~S.~Park}
\affiliation{Department of Physics, Kyungpook National University, Daegu 41566, Republic of Korea}
\author{K.~S.~Park}
\affiliation{Center for Underground Physics, Institute for Basic Science (IBS), Daejeon 34126, Republic of Korea}
\author{K.~Park}
\affiliation{Center for Underground Physics, Institute for Basic Science (IBS), Daejeon 34126, Republic of Korea}
\author{S.~D.~Park}
\affiliation{Department of Physics, Kyungpook National University, Daegu 41566, Republic of Korea}
\author{R.~L.~C.~Pitta}
\affiliation{Physics Institute, University of S\~{a}o Paulo, 05508-090, S\~{a}o Paulo, Brazil}
\author{H.~Prihtiadi}
\affiliation{Department of Physics, Universitas Negeri Malang, Malang 65145, Indonesia}
\author{C.~Rott}
\affiliation{Department of Physics and Astronomy, University of Utah, Salt Lake City, UT 84112, USA}
\author{K.~A.~Shin}
\affiliation{Center for Underground Physics, Institute for Basic Science (IBS), Daejeon 34126, Republic of Korea}
\author{D.~F.~F.~S. Cavalcante}
\affiliation{Physics Institute, University of S\~{a}o Paulo, 05508-090, S\~{a}o Paulo, Brazil}
\author{M.~K.~Son}
\affiliation{Department of Physics and Institute for Sciences of the Universe, Chungnam National University, Daejeon 34134, Republic of Korea}
\author{N.~J.~C.~Spooner}
\affiliation{Department of Physics and Astronomy, University of Sheffield, Sheffield S3 7RH, United Kingdom}
\author{L.~T.~Truc}
\affiliation{Department of Physics, Kyungpook National University, Daegu 41566, Republic of Korea}
\author{L.~Yang}
\affiliation{Department of Physics, University of California San Diego, La Jolla, CA 92093, USA}
\author{G.~H.~Yu}
\affiliation{Center for Underground Physics, Institute for Basic Science (IBS), Daejeon 34126, Republic of Korea}
\collaboration{COSINE-100 Collaboration}

\begin{abstract}
We report new constraints on the spin-dependent scattering cross section between low-mass dark matter and protons using data collected by the COSINE-100 experiment. By implementing a specialized event selection process using a multi-layer perceptron and robust noise mitigation, this analysis pioneers a detection threshold of 3 and 4 isolated peaks, corresponding to the reconstructed photoelectrons, which are significantly lower than the 8 photoelectron threshold used in previous analyses. In this unstudied few-photoelectron regime, where instrumental noise and phosphorescence are prevalent, we utilize a phenomenological background model to search for the annual modulation signal expected from the Standard Halo Model. No statistically significant annual modulation is observed in our data. We derive new 90\% confidence level (C.L.) upper limits for the spin-dependent DM--proton cross section, establishing the world's most stringent constraints in the $1.75–2.25\text{ GeV/c}^2$ mass range. Furthermore, by incorporating the Migdal effect, we extend the experimental sensitivity to the sub-GeV/c$^2$ regime, setting world-leading limits in the $15–58\text{ MeV/c}^2$ range. These results demonstrate the capability of NaI(Tl) target materials to probe previously unexplored regions of the dark matter parameter space. 
\end{abstract}

\maketitle

\emph{Introduction.}|The existence of Dark Matter (DM), constituting approximately $27\%$ of the Universe's total energy density, is strongly supported by a wide range of astrophysical and cosmological observations~\cite{Persic:1995ru,bertone2018,aghanim2020,Clowe:2006eq,Scognamiglio:2026phv}. Weakly Interacting Massive Particles (WIMPs) have long been considered among the most compelling candidates for DM~\cite{lee1977,schumann2019,Akerib:2022ort,Billard:2021uyg}. Direct detection experiments aim to observe the elastic scattering of WIMPs off atomic nuclei, resulting in nuclear recoils~\cite{goodman1985a, Akerib:2022ort,Billard:2021uyg,tasi2025}. Recently, increasing attention has focused on low-mass DM with masses below a few $\text{GeV/c}^{2}$, where the resulting small recoil energies require the use of advanced detectors with energy thresholds reduced below 1\,keV~\cite{supercdmscollaboration2018,abdelhameed2019,arora2025a,adari2025a}. 
%where the expected recoil energy appears too small to be observed below typical detector thresholds. 

The search for such light DM candidates can be addressed by considering the Migdal effect~\cite{migdal1939, migdal1941, ibe2018, vergados2020}. This effect occurs when the inherent resilience of atomic electrons is insufficient to accommodate the sudden acceleration of a recoiling nucleus, leading to excitation or ionization. This process allows experiments to extend their sensitivity into  the sub-GeV/c$^{2}$ mass range~\cite{xenoncollaboration2019, cosine-100collaboration2022a,agnes2023aaa,yu2025}. Recently, the Migdal effect has been observed in neutron--nucleus interactions with nuclear recoil energy above 35\,keV and electron energy of $5-10$\,keV~\cite{Yi:2026fmf}, where the observed effect is consistent with theoretical predictions.

The COSINE-100 experiment~\cite{adhikari2018b,adhikari2018a}, utilizing NaI(Tl) scintillating crystals, offers an advantage for probing spin-dependent (SD) DM--proton interactions due to the presence of $^{23}_{11}\text{Na}$ and $^{127}_{53}\text{I}$ isotopes, both of which are proton-odd nuclei with nearly 100\% natural abundance. Furthermore, the relatively low atomic mass of sodium enhances sensitivity to low-mass dark matter compared to experiments employing heavier target materials. This intrinsic sensitivity, combined with the development of techniques to reach the few-photoelectron (PE) regime, enables COSINE-100 to probe lower-mass SD DM--proton parameter space than was accessible in the previous analysis~\cite{yu2025}.
% This intrinsic sensitivity, combined with the development of techniques to reach the few-photoelectron (PE) regime, allows us to effectively bridge the gap in sub-GeV/c$^2$ dark matter searches.

In this Letter, we present the result of a search for low-mass DM using the COSINE-100 data. Crucially, we use a pioneering few-PE threshold, analyzing data at the 3 and 4 pulse peaks level. This threshold is significantly below the 8 PE threshold employed in previous COSINE-100 analyses~\cite{COSINE-100:2024wji,COSINE-100:2024ola,carlin2025a,yu2025}. 
We exploit this low-threshold data to search for an annual modulation signal and set new constraints on the SD DM-proton cross section, including limits extended by incorporating the Migdal effect.

\emph{Experimental setup.}|The COSINE-100 experiment situated at the Yangyang Underground Laboratory in South Korea was  protected by an overburden of approximately 700\,m of rock~\cite{prihtiadi2018,prihtiadi2020}. The detector commenced data acquisition in October 2016 and operated until March 2023, totaling 5.8 years of effective live time~\cite{carlin2025a}. The primary detection mass consists of eight low-background NaI(Tl) scintillating crystals with a combined mass of 106\,kg~\cite{adhikari2018b,adhikari2018a}.  Each crystal is coupled to two high-quantum-efficiency Hamamatsu R12669SEL 3-inch photomultiplier tubes (PMTs) at opposite ends. 

To achieve a low-energy threshold, we require a minimum trigger condition: the detection of at least one PE in each of the two PMTs within a 200\,ns coincidence time window~\cite{adhikari2018c}. This trigger condition efficiently acquires 3 and 4 pulse peaks events that are of interest in this analysis.
To further reduce backgrounds, the crystal array is immersed in a 2,200\,l liquid scintillator (LS) veto~\cite{adhikari2020b}, and the entire assembly is shielded by a nested structure comprising 3\,cm copper, 20\,cm lead, and an outer 3\,cm plastic scintillator panel array to veto cosmic muons~\cite{adhikari2018b}.

This analysis focuses on single-hit events because, for the small DM--nucleon scattering cross sections considered here, the probability of a DM particle scattering in more than one crystal is negligible. The single-hit events are defined as energy depositions occurring in only one crystal, with no coincident signals observed in the other seven crystals or the veto detectors. Multiple-hit events are utilized to understand and characterize background characteristics. Due to the high PMT noise rate of crystal 1 and the low light yields of crystals 5 and 8, we utilize only five crystals (crystals 2, 3, 4, 6, and 7), consistent with previous analyses~\cite{adhikari2018,carlin2025a, yu2025}. To avoid short-lived cosmogenic contributions~\cite{barbosadesouza2020,COSINE-100:2024ola}, approximately 2.2 years of the initial operation period was excluded, leaving a relatively stable 3.6 year dataset for the analysis.

\emph{Event selection.}|Throughout this Letter, `noise' denotes any recorded event that does not originate from a prompt scintillation signal of a single energy deposition in a crystal, encompassing instrumental noise (PMT dark current, after-pulses, coincident PMT events, and Cherenkov emission in the PMT glass) and phosphorescence--delayed light emission following earlier high-energy (HE) interactions. Events from radioactive decays are referred to as `radioactive background'.

\begin{figure}[t]
    \includegraphics[width=1.0\columnwidth]{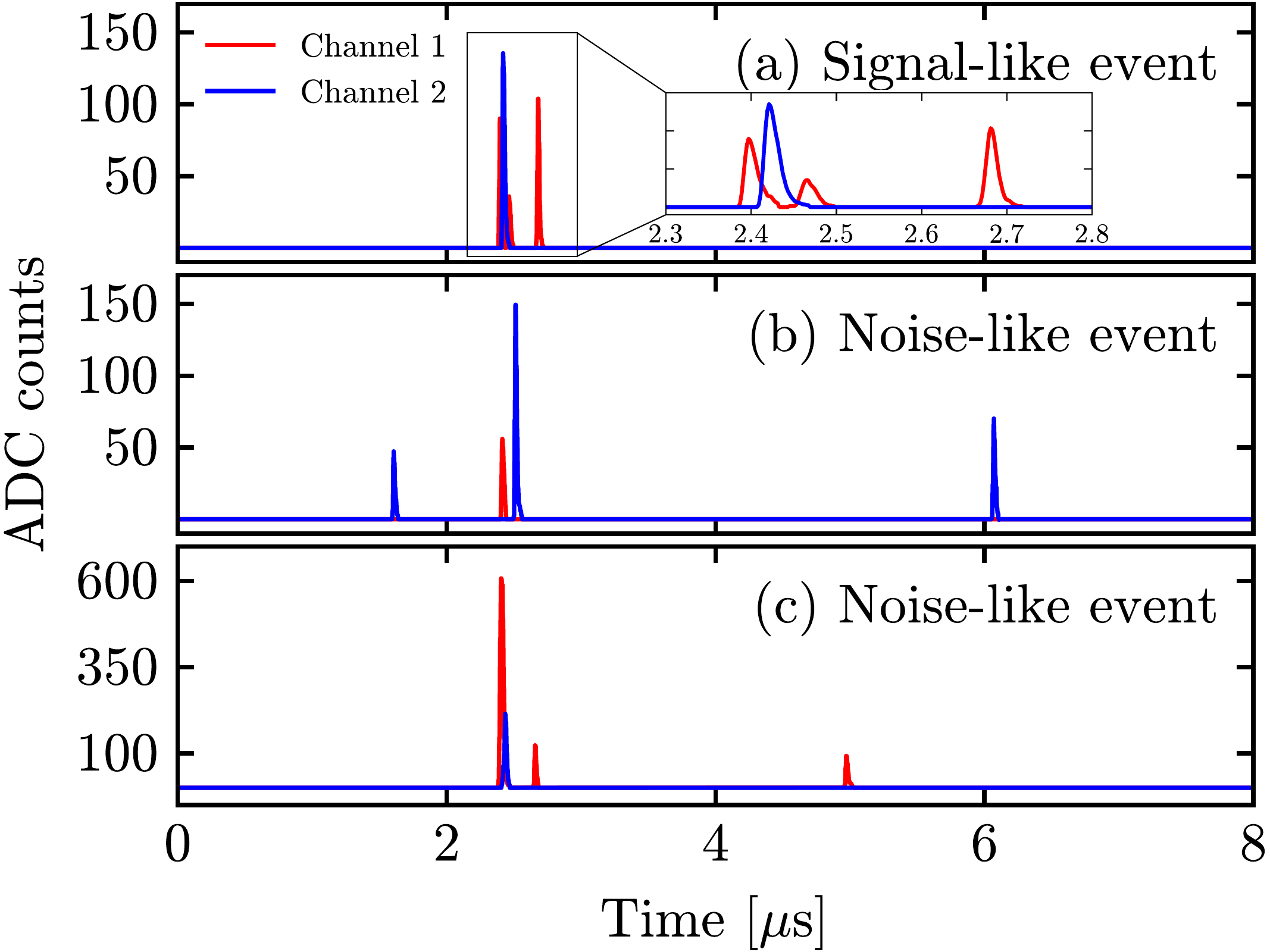}
    \caption{Representative waveforms  for NC-4 events in a NaI(Tl) crystal. (a) A signal-like event passing the MLP selection criteria, with clusters distributed within the characteristic 200\,ns scintillation decay time. The inset provides a zoomed-in view of the 2.3–2.8\,$\mu\text{s}$ region, clearly illustrating the individual photon clusters within the pulse. (b) A noise-like event rejected by the MLP selection due to clusters distributed over a broader time interval, characteristic of phosphorescence. (c) A noise-like event induced by Cherenkov radiation in the PMT glass, characterized by an anomalous single-cluster profile, rejected by the cluster charge cut.
    % (c) A noise-like event induced by Cherenkov radiation in the PMT glass, characterized by an anomalous single-cluster profile, also rejected by the MLP.
    }
    \label{fig:1}
\end{figure}

The exploration of low-mass DM requires pushing the detection threshold down to the few-PE level, which is substantially below previous COSINE-100 analysis thresholds. To target these few-PE energy depositions, we define the number of clusters (NCs) as the count of isolated, localized peaks in a waveform that exceed a pedestal-based threshold using the clustering algorithm described in Ref.~\cite{Kims:2005dol} and exhibit a single PE-like profile as shown in Fig.~\ref{fig:1}. Events with NC = 3 and NC = 4, hereafter denoted NC-3 and NC-4, are the targets of this analysis.

To characterize signal behavior at this level, we utilize a detailed NaI(Tl) waveform simulation~\cite{choi2024} that incorporates measured scintillation characteristics and single-PE pulse shapes. This simulation allows us to evaluate the detector's trigger efficiency by applying the hardware requirements directly to the simulated events.  We obtained a trigger efficiency of 62.5\% and 81.0\% for NC-3 and NC-4 events, respectively. The results confirm that events generated with a fixed number of PEs produce a reconstructed NC distribution centered around the input value, providing a reliable mapping between physical energy depositions and our observed NC counts. %In this study, we specifically focus on the 3 and 4-NC level to maximize sensitivity to the lowest DM masses.

A three-step event selection process was implemented to contend with the unique noise contributions in this unstudied region. These are dominated by coincident PMT-induced noise--principally Cherenkov emission in the PMT glass and phosphorescence following energetic events--which, although originating in a single PMT, is registered in both channels and is discriminated from genuine scintillation by its anomalous cluster timing and inter-PMT charge asymmetry (Fig.~\ref{fig:1} (b,c)).

First, a Deadtime cut was applied to abate phosphorescence events--delayed light emissions following high-energy precedent events--that can mimic genuine low-energy signals~\cite{cherwinka2016}. 
A deadtime of 0.3 s was imposed after HE events between 1 and 3.2 MeV, and 2.2 s after events above 3.2 MeV.
This initial filter achieved a high signal retention of 97\% while simultaneously reducing noise events by approximately 31.5\% in NC-3 and 22.2\% in NC-4.

Then, a complementary cluster charge cut was applied to remove abnormally large single clusters typically caused by Cherenkov radiation in the PMT glass. The cut boundary was defined conservatively to preserve a signal efficiency of approximately 97\%, ensuring negligible impact on genuine low-energy scintillation events. This cut reduces the noise level by approximately 10\% in  NC-3 and 15\% in NC-4.

To achieve further separation between signal and noise, a machine learning technique using a multi-layer perceptron (MLP) in the ROOT \texttt{TMVA} toolkit~\cite{brun1997,hoecker2009} was employed, similar to recent COSINE-100 analyses~\cite{COSINE-100:2024wji,COSINE-100:2024ola,yu2025}. The MLP was trained separately for each crystal and NC count using kinematic variables such as the time difference between clusters and the charge asymmetry between the two PMTs. The signal dataset for training was derived from the aforementioned waveform simulations, while the noise dataset was taken directly from the noise-dominated physics data. Since noise overwhelmingly dominates this sample, the small genuine-signal component it contains is classified on the signal side of the discriminator rather than biasing the noise rejection, so no purified noise selection is required.

The MLP training is designed to retain signal-like events, characterized by isolated clusters well-distributed within the 200\,ns decay time of NaI(Tl), as illustrated in Fig.~\ref{fig:1} (a). Conversely, it is trained to reject noise-like events where clusters are distributed over much wider windows typically caused by phosphorescence, as shown in Fig.~\ref{fig:1} (b). Given the difficulty of perfect separation, we chose selection criteria resulting in a 50\% signal efficiency, which provided an average noise rejection of approximately 91\% for NC-3 and 97\% for NC-4 events that already passed the Deadtime selection.
Fig.~\ref{fig:1} (c) illustrates the noise-like events rejected by the preceding cluster charge cut, such as those induced by Cherenkov radiation. 
% It is also identifies and rejects events exhibiting a large size single cluster typically caused by Cherenkov radiation in the PMT glass, as shown in Fig.~\ref{fig:1}(c).

Only single-hit events passing the Deadtime, cluster charge, and MLP cuts were used in the subsequent annual modulation search. After applying all selection criteria, the stability of the detector was monitored using the multiple-hit events sideband. Large rate fluctuations were observed in crystal 2 for the initial 3.6 years of the data period. To maintain data integrity, this unstable period for crystal 2 was excluded from the analysis. Consequently, the final dataset utilized for the search consists of the latest stable 3.6 year period for crystals 3, 4, 6, and 7, and a 2 year period for crystal 2, corresponding to an effective exposure of 286\,kg$\cdot$yr.

\emph{Annual modulation.}|Following the rigorous event selection using Deadtime, cluster charge, and MLP cuts, the remaining dataset consists of a mixture of radioactive backgrounds and unexpected noise contributions from PMT-induced events and phosphorescence, and the potential signal from DM interactions. Because a comprehensive physics-based background simulation is currently unfeasible for the few-PE region, we adopt a phenomenological approach to model the time-dependent background and search for a superimposed annual modulation signal. The time evolution of the event rate, $F(t;A)$, is modeled using a function that incorporates a constant baseline $C$, an exponential component to account for decaying background elements, and a sinusoidal function representing the DM-induced modulation:
\begin{equation}
  F(t;A) = A\cos\left[\frac{2\pi}{T}(t-\phi)\right] + \exp(p_0 + p_1 t) + C,
  \label{eq:model}
\end{equation}
where $A$ is the modulation amplitude, $p_{0}$ and $p_{1}$ are the parameters for exponential background. 
% In the annual modulation fit, the period is fixed to \(T=365.25\) days and the phase to the standard expectation of maximum rate around June 2. These choices are motivated by the generic annual variation of the relative velocity between the Earth-based detector and the Galactic dark-matter halo, and do not require the full set of Standard Halo Model assumptions. The SHM assumptions enter in the subsequent model-dependent calculation of the expected recoil spectrum and modulation amplitude used to derive limits.

The period $T$ and phase $\phi$ are fixed to 365.25\,days~(1\,year) and 152.5\,days, respectively, as these follow directly from the Earth's orbital motion around the Sun and are consistent with the expectation of the Standard Halo Model~\cite{lewin1996, freese1988, freese2013}. 
Simultaneous fits across the crystals are performed on 15-day time-binned data for the stable period. 
To rigorously evaluate the stability of the signal region, we utilize a sideband consisting of events neighboring the MLP selection region. This sideband includes events classified as noise by the 50\% signal acceptance cut but contains a comparable number of events. According to waveform simulations, approximately 20\% and 17\% of genuine signal events for the NC-3 and NC-4 categories, respectively, are accepted into the noise-classified sideband region.  
%4-year period from crystals 3, 4, 6, and 7 and 2-year period from crystal 2. 

\begin{figure*}[t]
  \centering
  \begin{minipage}[b]{0.48\textwidth}
    \centering
    \includegraphics[width=\linewidth]{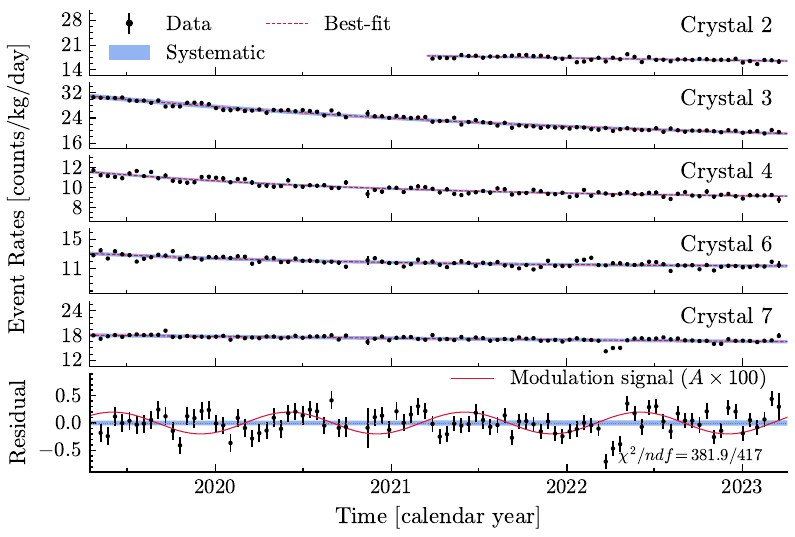}
  \end{minipage}\hfill
  \begin{minipage}[b]{0.48\textwidth}
    \centering
    \includegraphics[width=\linewidth]{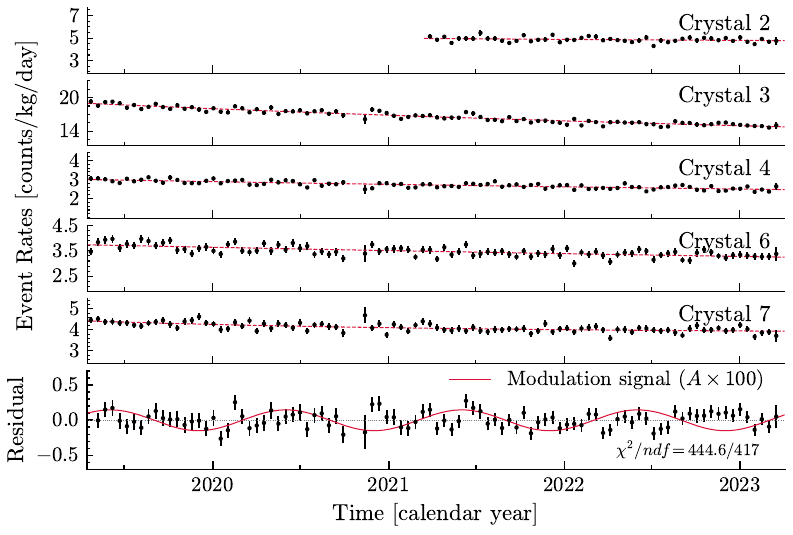}
  \end{minipage}
  \caption{Event rates as a function of time for the five NaI(Tl) crystals used in the analysis, overlaid with the best-fit annual modulation model for single-hit events. Data are grouped in 15-day bins, and the fits are performed simultaneously across all crystals with a fixed phase of 152.5 days and a shared modulation amplitude $A$. The bottom panels show the residuals from the fits. The residuals are defined as the data minus the full best-fit model of Eq.~\eqref{eq:model}, including both the background and cosine modulation terms. (Left) The NC-3 dataset is shown with the blue shaded bands representing the systematic uncertainties ($\delta u_i$). (Right) The NC-4 dataset is shown.}
  \label{fig:2}
\end{figure*}

We observed that the NC-3 dataset exhibited significant fluctuations, resulting in a reduced $\chi^2$ greater than unity. To account for this instability in the signal region, we derived a scale factor for each crystal, specifically calculated to bring the reduced $\chi^2$ of the sideband to unity. This value is then assigned as a relative crystal-specific systematic uncertainty ($\delta u_i$), derived from the sideband stability analysis. The $\chi^2$ function is defined as: 
\begin{equation}
    \chi^2=\sum_i^{N_\mathrm{crystal}}\sum_j^{N_\mathrm{time}}
    \frac{\left[M_{ij}-F_i(t_j;A)\right]^2}{M_{ij}+\delta u_{i}^2M_{ij}^2},
\end{equation}
where $M_{ij}$ is the observed count in time bin $j$ of crystal $i$ and $F_i(t_j;A)$ is the model in Eq.~\eqref{eq:model}. The systematic uncertainties are treated as fully uncorrelated between bins. Because both the crystal 2 exclusion and the $\delta u_{i}$ factors are derived from sidebands independent of the single-hit modulation amplitude, these data-driven choices do not bias the result.

\begin{table}[t]
    \centering
    \begin{tabular}{c cc}
        \toprule
        \multirow{2}{*}{Number of clusters [NCs]} & 
        \multicolumn{2}{c}{Amplitude ($A$) [counts/kg/day]} \\
        \cmidrule(lr){2-3}
         & Single-hit & Multiple-hit \\
        \midrule
        3 & $-0.002 \pm 0.026$ & $0.002 \pm 0.007$ \\
        4 & $0.001 \pm 0.011$ & $0.002 \pm 0.003$ \\
        \bottomrule
    \end{tabular}
    \caption{Best-fit annual modulation amplitudes ($A$) obtained from the simultaneous fit of five crystals, performed separately for the NC-3 and NC-4 pulse cluster bins. Results for both single-hit (signal) and multiple-hit (background control) events are consistent with the null hypothesis, showing no statistically significant modulation.}
    \label{tab:amplitude}
\end{table}

Fig.~\ref{fig:2} presents the results of simultaneous fits performed across the five crystals, conducted separately for the NC-3 (Left) and NC-4 (Right) datasets. Goodness-of-fit is quantified by $\chi^2/\text{NDF}$ (number of degrees of freedom), with the observed modulation amplitude $A$ amplified by 100 times for visibility, as indicated in each last panel. Given the sufficiently large NDF, reduced $\chi^2$ values close to unity are observed, proving statistically consistent fits.
%For completeness, the goodness-of-fit can also be interpreted in terms of the corresponding p-value derived from the $\chi^2$ distribution for the given NDF.}

% , where 89.7\% in 3-NC and 17.8\% in 4-NC data fitting results.}
No statistically significant modulation was observed in either case. Identical fits performed on multiple-hit events consistently showed no modulation, as summarized in Table~\ref{tab:amplitude}. In the absence of a modulated signal, these results are used to establish new constraints on the SD DM--proton cross section.

\emph{Dark matter limits.}|To translate the null result from the annual modulation search into constraints on DM cross sections, we calculate the expected signal based on the Standard Halo Model. The differential recoil rate induced by DM–nucleon scattering is calculated using the {\sc dmdd} package~\cite{gluscevic2015, gluscevic2015a,anand2014}. 
For the Migdal channel, we adopt the atomic ionization probabilities of Ref.~\cite{ibe2018}, computed in the isolated-atom approximation; solid-state corrections to this approximation are expected only at energy transfers of order tens of eV~\cite{Essig2020Migdal,Knapen2021Migdal}, well below our approximately 0.2\,keV analysis threshold. 
The expected modulation amplitude, $A'(\sigma_{\mathrm{DM}})$, for each DM model is derived from the difference in recoil rates between the maximum and minimum Earth speeds ($v_{\mathrm{max}} \approx 247$ km/s and $v_{\mathrm{min}} \approx 217$ km/s), reflecting the annual variation of the relative velocity between the Earth-based detector and DM particles in the Galactic halo~\cite{goodman1985a,lewin1996,freese2013,schumann2019}. These physical spectra are converted to observable NC distributions through a comprehensive detector response model, applying measured light yields~\cite{adhikari2018b} and quenching factors~\cite{lee2024}, processed through detailed waveform simulations~\cite{choi2024}.

Because the expected DM signal provides a linked constraint across both the NC-3 and NC-4 datasets, we reconstruct the $\chi^2$ function to simultaneously fit all selected crystals and NC bins:
\begin{equation}
    \chi^2(\sigma_{\textrm DM})=\sum_i^\mathrm{NC}\sum_j^{N_\mathrm{crystal}}\sum_k^{N_\mathrm{time}}
    \frac{\left[M_{ijk}-F_{ij}(t_k;A')\right]^2}{M_{ijk}+\delta u_{ij}^2M_{ijk}^2}
    +\left(\frac{\alpha_{ij}}{\delta c_{ij}}\right)^2,
    \label{eq:chi2p}
\end{equation}
where $M_{ijk}$ is the observed count in time bin $k$ for crystal $j$ and NC bin $i$, and $F_{ij}(t_k; A')$ is the DM-induced modulation model as defined in Eq.~\eqref{eq:model}, where the modulation amplitude $A$ is replaced by the expected amplitude $A'$. In this framework, $A'$ is directly tied to the DM cross section: 
\begin{equation}
    A'(\sigma_{\textrm DM})=(1+\alpha_{ij})f_{ij}(m_\mathrm{DM})\,\sigma_\mathrm{DM},
\end{equation}
where $f_{ij}(m_\mathrm{DM})$ is a conversion factor from the cross-section $\sigma_\mathrm{DM}$ to event rates (counts/kg/day), and $\alpha_{ij}$ is a nuisance parameter accounting for light yield and quenching factor uncertainties with a relative uncertainty $\delta c_{ij}$. 
The quenching factors are constrained to approximately 4\% by the modified Lindhard-model fit to our measurements~\cite{lee2024}, propagating to less than 3.5\% on the 90\% C.L. limit, whereas the dominant systematic is the crystal-specific noise-rate instability $\delta u_{ij}$ (Eq.~\eqref{eq:chi2p}), which contributes up to ~30\%, with the light yield term remaining subdominant. 
Upper limits are established at a 90\% C.L. based on maximum likelihood estimation using Eq.~\eqref{eq:chi2p}. A Probability Density Function (PDF) for each mass is defined as: 
\begin{equation}
\text{PDF} = C \exp \left( -\frac{\Delta\chi^2}{2} \right),\label{eq:pdf}
\end{equation}
where $C$ is a normalization factor ensuring the PDF integrates to unity over the physical region ($\sigma_{\textrm DM} \geq 0$), and $\Delta\chi^2 = \chi^2(\sigma_{\textrm DM}) - \chi^2_{\text{min}}$. In this analysis, $\chi^2_\text{min}$ denotes the minimum value of $\chi^2(\sigma_\text{DM})$.

\begin{figure}[t]
  \centering
  \begin{minipage}[b]{0.45\textwidth}
    \centering
    \includegraphics[width=\linewidth]{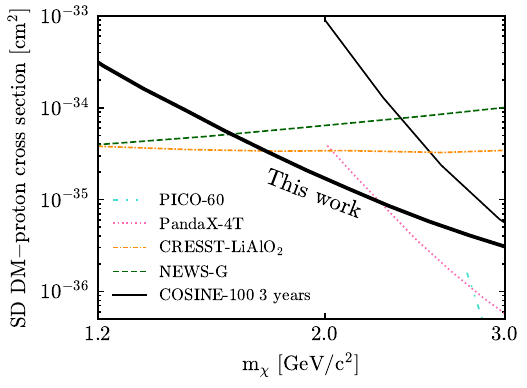}
  \end{minipage}\hfill
  \begin{minipage}[b]{0.46\textwidth}
    \centering
    \includegraphics[width=\linewidth]{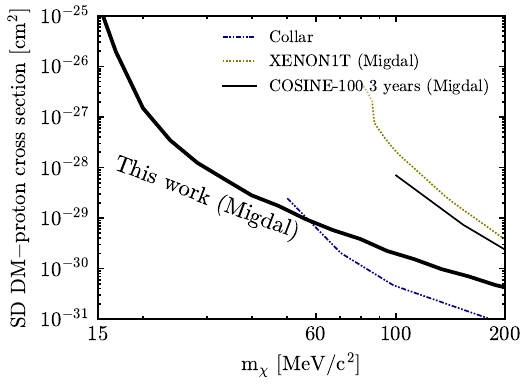}
  \end{minipage}
  \caption{The derived 90\% C.L. upper limits on the SD DM-proton cross section as a function of DM mass m$_\chi$. The thick solid line represents the result of this work. (Top) Constraints in the $1.2–3.0$\,GeV/c$^2$ mass region for standard SD nuclear recoils. Our result provides new constraints in the $1.75–2.25$\,GeV/c$^2$ range, a region not previously explored by the compared experiments: PICO~\cite{picocollaboration2019}, PandaX-4T~\cite{zhang2025a}, CRESST-$\text{LiAlO}_2$~\cite{angloher2022}, NEWS-G~\cite{arora2025a}, and the recent COSINE-100 3\,years data result ~\cite{yu2025}. (Bottom) Extended sensitivity in the $15–200$\,MeV/c$^2$ region enabled by the Migdal effect. Our results establish stringent new limits in the $15–58$\,MeV/c$^2$ range, significantly extending the reach beyond the 100\,MeV/c$^2$ limit achieved in previous COSINE-100 3\,years data result~\cite{yu2025}. This work surpasses the previous experimental reach of Collar~\cite{collar2018a} and XENON1T~\cite{xenoncollaboration2019}. }
  % In both panels, the results of this work (thick solid lines) are compared with previous COSINE-100 results (thin solid lines) and other leading experimental constraints (dashed and dotted lines).}
  \label{fig:3}
\end{figure}

As shown in Fig.~\ref{fig:3} (Top), we set new constraints on the SD DM--proton cross section in the $1.2–3.0$\,GeV/c$^2$ mass region. Notably, this work probes a previously unexplored parameter space between 1.75 and 2.25 GeV/c$^2$ that has not been constrained by other experiments~\cite{picocollaboration2019, zhang2025a, angloher2022, yu2025,arora2025a, xenoncollaboration2019, collar2018a}. Furthermore, the inclusion of the Migdal effect extends our reach to masses as low as 15\,MeV/c$^2$, as illustrated in Fig.~\ref{fig:3} (Bottom). While our previous analysis (COSINE-100 3\,years) using an 8\,PE threshold reached limits down to 100\,MeV/c$^2$~\cite{yu2025}, this NC-3 and NC-4 threshold analysis establishes stringent new limits between 15 and 58\,MeV/c$^2$. These results demonstrate the capability of NaI(Tl) target materials to probe previously unexplored regions of the dark matter parameter space.

\emph{Conclusion.}|We have established a new detection threshold for NaI(Tl) at the 3-PE level. This low-threshold analysis successfully isolates signals from PMT noise and phosphorescence, allowing us to probe unexplored parameter spaces. We established new 90\% C.L. constraints for SD DM--proton cross section in the $1.75–2.25$\,GeV/c$^2$ range and, via the Migdal effect, in the $15–58$\,MeV/c$^2$ mass range. This provides a critical foundation for COSINE-100U~\cite{Lee:2024wzd} at Yemilab, where higher light yields will further enhance sensitivity to light dark matter.

\vspace{1em}

We thank the Korea Hydro and Nuclear Power (KHNP) Company for providing underground laboratory space at Yangyang and the IBS Research Solution Center (RSC) for providing high performance computing resources. 
This work is supported by:  the Institute for Basic Science (IBS) under project code IBS-R016-A1,  NRF-2021R1A2C3010989, NRF-2021R1A2C1013761, RS-2024-00356960, RS-2025-25442707 and RS-2025-16064659, Republic of Korea;
NSF Grants No. PHY-1913742, United States; 
STFC Grant ST/N000277/1 and ST/K001337/1, United Kingdom;
Grant No. 2021/06743-1, 2022/12002-7, 2022/13293-5 and 2025/01639-2 FAPESP, CAPES Finance Code 001, CNPq 304658/2023-5, Brazil;
UM Internal Grant non-APBN 2025, Indonesia.

% \vspace{1em}
% \section{Appendix}\textbf{Spin-independent} limit on WIMP-nucleon interaction is ..

\bibliography{zotero_dm}

\end{document}